\documentclass[amsmath, amssymb, english, aps, prl, twocolumn, reprint,floatfix, superscriptaddress,showpacs]{revtex4}
\usepackage[english]{babel}
\usepackage{graphicx}
\usepackage{verbatim}
\def \sriro{Sr${}_2$IrO$_{4}$}
\def \srrho{Sr${}_2$RhO$_{4}$}
\def \srruo{Sr${}_2$RuO$_{4}$}
\def \t2g{t$_{2g}$}
\def \eg{e$_{g}$}
\def \dxy{d$_{xy}$}

\def \dx2y2{d$_{x^2-y^2}$}

\def \LDAU{LDA+U}

\def \LDADMFT{LDA+DMFT}

\def \TMO{TMO}

\def \cRPA{cRPA}
\def \ARPES{ARPES}
\def \j32{$j_{\textrm{eff}}$=$3/2$}
\def \jeff12{$j_{\textrm{eff}}$=$1/2$}

\def\expect#1{\mathinner{\langle{#1}\rangle}}

{\catcode`\|=\active
  \gdef\expect#1{\left<\mathcode`\|"8000\let|\bravert {#1}\right>}}
\def\bravert{\egroup\,\vrule\,\bgroup}

\begin{document}

\title{Reduced effective spin-orbital
degeneracy and spin-orbital ordering in paramagnetic
transition metal oxides: Sr$_2$IrO$_4$ vs. Sr$_2$RhO$_4$}

\author{Cyril Martins}
\affiliation{Centre de Physique Th\'eorique, Ecole Polytechnique, CNRS,
91128 Palaiseau Cedex, France}
\affiliation{Japan Science and Technology Agency, CREST, Kawaguchi
  332-0012, Japan}
\author{Markus Aichhorn}
\affiliation{Institute of Theoretical and Computational Physics, TU Graz, Petersgasse 16, Graz, Austria}
\affiliation{Centre de Physique Th\'eorique, Ecole Polytechnique, CNRS,
91128 Palaiseau Cedex, France}
\author{Lo\"ig Vaugier}
\affiliation{Centre de Physique Th\'eorique, Ecole Polytechnique, CNRS,
91128 Palaiseau Cedex, France}
\author{Silke Biermann}
\affiliation{Centre de Physique Th\'eorique, Ecole Polytechnique, CNRS,
91128 Palaiseau Cedex, France}
\affiliation{Japan Science and Technology Agency, CREST, Kawaguchi
  332-0012, Japan}

\begin{abstract}
We discuss the notions of
spin-orbital polarization and ordering
in paramagnetic materials,
and address their consequences in
transition metal oxides.
Extending the combined density 
functional and dynamical mean field theory 
scheme to the case of materials 
with large spin-orbit interactions,
we investigate the electronic excitations
of the paramagnetic phases of Sr$_2$IrO$_4$ and Sr$_2$RhO$_4$.
We show that the interplay of spin-orbit interactions,
structural distortions and Coulomb interactions
suppresses spin-orbital
fluctuations. As a result, the room temperature
phase of Sr$_2$IrO$_4$ is a paramagnetic spin-orbitally
ordered Mott insulator.
In Sr$_2$RhO$_4$, the effective spin-orbital degeneracy
is reduced, but the material 
remains metallic, due to both,
smaller spin-orbit and
smaller Coulomb interactions. 
We find excellent agreement of our {\em ab-initio}
calculations for Sr$_2$RhO$_4$ with
angle-resolved photoemission, and make predictions for
spectra of the paramagnetic phase of Sr$_2$IrO$_4$.
\end{abstract}


\maketitle

Probably the most important consequence of relativistic
quantum mechanics in solids is the coupling of spin and
orbital degrees of freedom. The concept of spin-orbit  
coupling (SOC) has been known
for more than half a century, and its importance in magnetic
materials has been recognized early on. 
Nevertheless, SOC in non-magnetic materials has only recently 
become a hot topic in condensed matter physics, in the context of
topological insulators \cite{KanePRL95-2005}, 
within the search for systems for quantum computing applications. 
Still, the interplay of spin-orbit (SO) interactions with electronic
Coulomb interactions in paramagnetic materials
remains a largely unexplored field. This is at least partially due
to the seemingly mutually exclusive regimes of their acting:
SO interactions are strongest in heavy elements,
that is in transition metal compounds with $5d$ (and, to a lesser 
extent $4d$) electrons.
Coulomb interactions, on the other hand, are believed to be most 
efficient in $3d$ compounds, due to the more localized 
$3d$ orbitals. It is only recently that this separation has been questioned,
e.g. in \cite{MoonPRL101-2008, KimPRL101-2008,
  KimScience323-2009,PesinNP-2010}. 

In this Letter, we present a particularly striking example for the interplay 
of SOC and Coulomb interactions: in the room-temperature phase of 
the $5d$ transition metal oxide (TMO) \sriro\ the combined 
effect of SOC 
and distortions from the ideal K$_2$NiF$_4$-structure is strong enough
to suppress the effective
degeneracy of the coupled spin-orbital degrees of freedom. 
Even moderate
Coulomb interactions then induce a Mott localized state. The resulting
state -- despite of being paramagnetic and paraorbital, i.e. in the absence 
of any magnetic or orbital order -- displays a ``spin-orbital order'' 
in the sense that the only hole in the t$_{2g}$ manifold 
occupies a state of well-defined t$_{2g}$-projected total angular 
momentum $J^2$.
This notion generalizes the concept
of orbital order, well-known in TMOs, 
to the case where neither spin nor orbitals are 
good quantum numbers.
We present -- entirely from first-principle calculations -- a scenario
for the paramagnetic insulating state of \sriro, in comparison 
to the isostructural and isoelectronic $4d$ compound \srrho.
For the latter, we find
that the smaller SOC and Coulomb interactions induce a
partially spin-orbitally polarized metallic state. We calculate 
photoemission spectra for both materials and find excellent 
agreement with available experiments.

\sriro\ is a $5d$-\TMO\ with a tetragonal crystal structure 
whose symmetry is lowered from 
the K$_2$NiF$_4$-type, well-known in \srruo\ or La$_2$CuO$_4$,
by 11$^\circ$ rotation of its IrO$_6$ octahedra around 
the \textbf{c}-axis
as in its $4d$-counterpart 
\srrho\ \cite{HuangJSSC112-1994, ItohJSSC118-1995}. Although each Ir site 
accomodates $5$ electrons, 
\sriro\ exhibits an insulating behavior at all temperatures 
with an optical gap of 
about $0.26$~eV at room temperature \cite{MoonPRB80-2009}. Below $240$\,K, 
canted-antiferromagnetic (AF) order sets in, with an effective paramagnetic
moment of $0.5$\,$\mu_B$/Ir and a saturation moment 
of $0.14$\,$\mu_B$/Ir \cite{CaoPRB57-1998}. 
This phase has triggered much experimental and theoretical work
recently \cite{KimPRL101-2008, KimScience323-2009, JinPRB80-2009, WatanabePRL105-2010}, 
highlighting in particular the importance of the SOC $\zeta_{SO}$.
SOC was also shown to be relevant in \srrho, which is a 
paramagnetic metal down to $36$\,mK \cite{MoonPRB74-2006}:
density functional calculations within the local density 
approximation (LDA), augmented by Coulomb interactions within the \LDAU\ method,
reproduce the Fermi surface only if SO interactions are taken into account 
\cite{HaverkortPRL101-2008, LiuPRL101-2008}. 

\begin{figure}
 \begin{center}
  \includegraphics[width=0.78\linewidth,keepaspectratio]{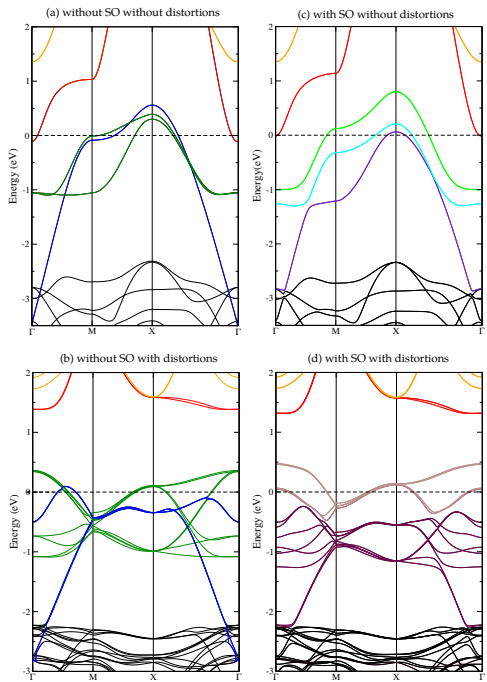}
 \end{center}
 \caption{\label{fig2}
 LDA band structures of \sriro\ without distortions and 
without SOC (a), 
 with distortions without SOC (b), 
without distortions with SOC (c) and 
 with both, distortions and SOC (d).}
\end{figure}

Here, 
we focus on the {\it paramagnetic} insulating phase of \sriro\ above $240$~K, 
which has not been addressed by theory before. 
We analyze the electronic properties by a combination of
the LDA with dynamical mean field theory (DMFT).
Our method is a generalization of the LDA+DMFT scheme as 
implemented in \cite{AichhornPRB80-2009}
based on 
the Wien2k package \cite{ReferenceWien2k}, extended 
to include SO interactions \cite{Martins-tbp}. 
The Coulomb interactions 
are calculated from the constrained random phase approximation (cRPA)
\cite{FerdiPRB70-2004}, taking matrix elements in the same set
of localized orbitals as used in the DMFT calculations
\cite{Vaugier-tbp}.
For \sriro\ we estimate the Slater integrals as $F^0$=$2.2$\,eV and
$J$=$0.3$\,eV, 
giving an intraorbital Hubbard U in the half-filled 
\jeff12\ orbitals of $U$=$2.25$\,eV \cite{footnote1}.

\begin{figure}
 \begin{center}
  \includegraphics[width=0.9\linewidth,keepaspectratio]{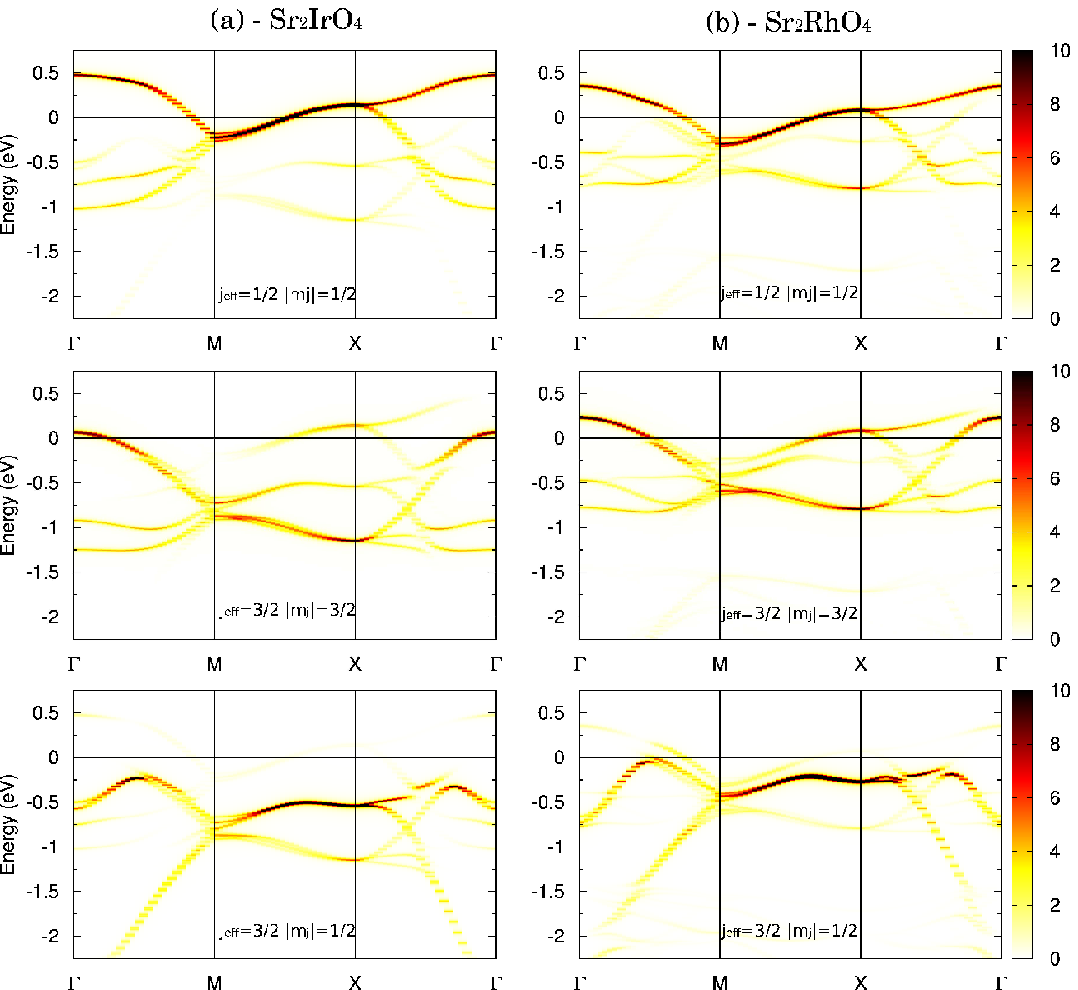}
 \end{center}
 \caption{\label{fig1}
LDA band structure of \sriro\ (left) and \srrho\ (right), projected
on the \jeff12 
 (top), \j32\ $|m_j|$=$3/2$ (middle), and \j32\ $|m_j|$=$1/2$ (bottom)
spin-orbitals.} 
\end{figure}

Within an LDA band picture, Fig.~\ref{fig2}~(d),
a metallic solution -- at variance with experiments -- is obtained for
\sriro. 
We construct Wannier functions for the t$_{2g}$ orbitals from the LDA
bands within the energy window $[-3.0,0.5]$\,eV. 
The large SOC of about $\zeta_{\textrm{SO}}$$\approx$$0.4$\,eV
splits these states into a quartet of 
states, commonly labeled as \j32, 
and a higher lying doublet \jeff12. 
Each state is twice degenerate in $\pm m_j$.
The four \j32\ states are almost filled;
we find $n_{3/2,|1/2|}$=$1.98$ and $n_{3/2,|3/2|}$=$1.84$ for the
\j32\ orbitals. The \jeff12\ states thus slightly exceed half-filling  
with $n_{1/2}$=$1.16$. In the left panel of Fig.~\ref{fig1}, we plot
the $j_{\textrm{eff}}$ character of the Wannier orbitals. The
four bands that cross the Fermi level are not purely formed by the 
\jeff12\ orbitals, but there is a slight overlap 
of the \jeff12\ and \j32\ $|m_j|$=$3/2$ characters, particularly
obvious at the $\Gamma$ point. Similar conclusions were drawn for the AF 
phase in Ref.~\cite{WatanabePRL105-2010}.

We now redefine the \jeff12\ and \j32\ $|m_j|$=$1/2$ atomic states, 
yielding a diagonal density matrix of the local problem. 
Thus, our definition takes into account 
the structural distortions and a tetragonal crystal field in each
IrO$_6$ octahedra. Our coefficients for the \jeff12\ state are
similar to those obtained for the AF phase in Ref.~\cite{JinPRB80-2009}.

\begin{figure}
 \begin{center}
  \includegraphics[width=0.8\linewidth,keepaspectratio]{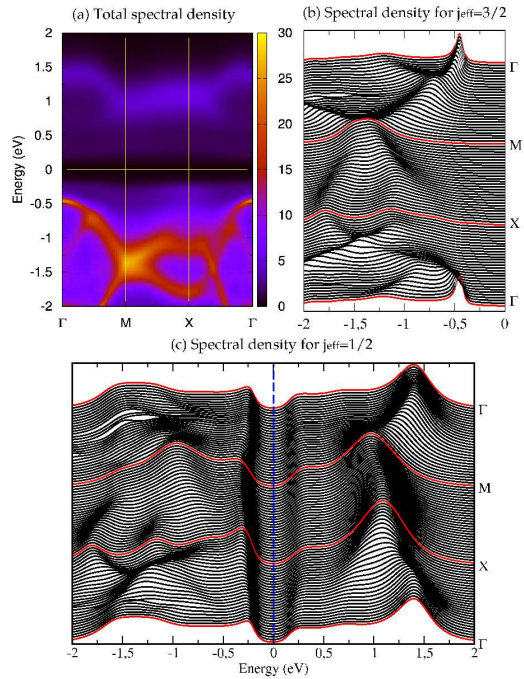}
 \end{center}
 \caption{\label{fig3} 
 Momentum-resolved spectral functions $A(\mathbf{k},\omega)$ 
 of the paramagnetic phase of \sriro\ from \LDADMFT\ at $T$=$300$\,K.  
}
\end{figure}

We now turn to our LDA+DMFT results. In
Fig.~\ref{fig3}, we display the corresponding spectral functions.
Within LDA+DMFT, supplemented by the \cRPA\ 
interactions, an insulating solution
with a Mott gap of the size of the optical gap measured at room
temperature (about $0.26$~eV \cite{MoonPRB80-2009}) is obtained. 
Despite the fact that the \ARPES\ measurements of
Ref.~\cite{KimPRL101-2008} were performed in the AF 
phase, a comparison of the total spectral function to the experimental
energy distribution curves shows qualitative and even quantitative 
agreement along the direction $\Gamma$-$X$ and $\Gamma$-$M$ (between
-$0.5$\,eV and -$1.5$\,eV). Along the $M$-$X$ direction 
a band around -$1$\,eV is identified. From the
orbitally-resolved spectral functions (Fig.~\ref{fig3} b and c), one
can locate the lower \jeff12\ Hubbard band at about
-$0.5$\,eV, in agreement with the \ARPES\ data. 

The spin-orbital polarization is enhanced when taking into account 
Coulomb correlations: the Wannier orbital \jeff12\ is now exactly
half-filled, and the upper Hubbard band is of \jeff12\ type only. 
We are thus dealing with a state exhibiting a ``spin-orbital
order'' in the sense of a well-defined $j_{\textrm{eff}}$ quantum number.
In contrast to the AF phase, where it is also the \jeff12\ state that
carries the hole, we do {\em not} have an ordering of the
corresponding $m_{j_\textrm{eff}}$=$\pm 1/2$ quantum number. 
Also, the orbital occupations of the original
t$_{2g}$ orbitals to the \jeff12\ spin-orbital are roughly equal. 
We are thus facing the remarkable situation of ``spin-orbital
order'', with {\em neither orbital nor magnetic order}.

As it is well-known from studies of multi-orbital Hubbard models
the critical interaction for the formation of the insulating
state is lowest for the one-band case, and increases 
with degeneracy (for a review see \cite{GeorgesJdeP-2004}).
In \sriro, the Hubbard interactions calculated from \cRPA\
are large enough to induce a Mott insulating state in a 
half-filled one-orbital
but not in a quarter-filled two-orbital
or a 1/6-filled three-orbital system. 
We can thus conclude that the reduced effective spin-orbital
degeneracy is the reason for the insulating nature of \sriro.
An analogous suppression of {\em orbital} degeneracy, albeit purely 
induced by cristal field splittings, has been studied in the model
context in \cite{ManiniPRB66-2002, PoteryaevPRB78-2008}, 
and has been found to make
LaTiO$_3$ and
YTiO$_3$ Mott-insulating \cite{PavariniPRL92-2004}. 

In \sriro\ the suppression of {\em spin-orbital} fluctuations
is a consequence of the combined effect of SOC
and the tetragonal distortions. 
To demonstrate this, we do a numerical experiment
in which we switch on or off the two effects separately and
monitor the critical value of $U$ for the Mott
metal-insulator transition.

In Fig.~\ref{fig2}~(a) both the SOC and the structural
distortions are omitted (we use the lattice parameters of
Ref.~\cite{RandallJACS79-1957}). 
The band structure of \sriro\ is then very similar to the one of 
\srruo, except for the increased filling compared to the $d^4$ Ru-compound.
The t$_{2g}$ orbitals almost equally accomodate the $5$ 
electrons ($n_{d_{xy}}$=$1.612$, $n_{d_{xz},d_{yz}}$=$1.696$). 
The insulating Mott state is reached for $F^0$=$3.6$\,eV and 
$J$=$0.3$\,eV with enhanced orbital polarization.
The \dxy\ band gets filled with increasing correlations, approaching  
a completely filled \dxy\ band and two $3/4$-filled d$_{xz/yz}$ bands (which is
the atomic ground state).

In Fig.~\ref{fig2}~(b), the SOC is neglected but the
structural distortions are included. This lowers the symmetry of
\sriro\ and results in a four-times larger unit cell and four-folded
bands. Similarly as in \srrho\ \cite{HaverkortPRL101-2008}, 
a \dxy-\dx2y2\ hybridization gap opens between the \eg\ and
the t$_{2g}$ bands (between $0.4$ and $1.3$\,eV), and 
the \dxy\ band becomes almost filled. The orbital
polarization is thus enhanced compared to the undistorted case, 
and the Mott transition occurs for smaller values of $F^0$. 
Indeed, \LDADMFT\ gives the insulating state for $F^0$=$3.0$\,eV with
$J$=$0.3$\,eV. 

In Fig.~\ref{fig2}~(c), we show the band structure of ``undistorted''
\sriro\ but with SOC. Three $j_{\textrm{eff}}$ bands can be 
identified: the \jeff12\ one with an occupation of $n_{1/2}$=$1.20$
lies above the two \j32\ ones, with fillings $n_{3/2,|3/2|}$=$1.92$
and $n_{3/2,|1/2|}$=$1.96$. We find this three band system 
to be insulating at $F^0$=$3.0$\,eV with $J$=$0.3$\,eV. 
Although the problem is effectively reduced to an almost
half-filled one-band model, the large bandwidth of the
\jeff12\ band prevents a smaller value for the critical $F^0$. 

Finally, panel Fig.~\ref{fig2}-(d) depicts the band structure of 
the true compound. In this case, the Mott transition occurs   
between $F^0$=$2.1$ and $2.2$\,eV for $J=0.3$\,eV, the values that have
been estimated by our cRPA calculations. 
The combination of both the structural distortions
and the SOC is thus necessary for \sriro\ to be insulating.

 \begin{figure}
 \begin{center}
 \includegraphics[width=0.8\linewidth,keepaspectratio]{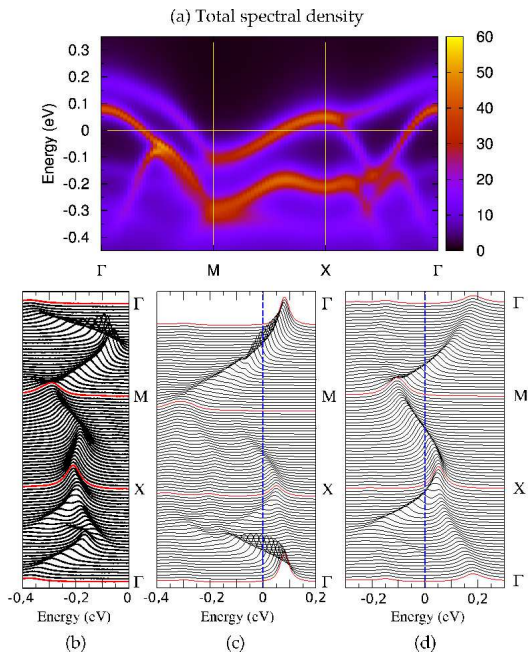}
 \end{center}
 \caption{\label{fig4} 
   Momentum-resolved spectral function $A(\mathbf{k},\omega)$ of \srrho\ (a). 
   Bottom pannels display the orbitally-resolved spectral 
   densities for \j32\ $|m_j|$=$1/2$ (b), 
   \j32\ $|m_j|$=$3/2$ (c) and \jeff12\ (d) at $T$=$300$\,K. 
} 
\end{figure}

In particular, we would expect \sriro\ to be metallic, if either the
distortions or the SOC were smaller. 
Such a situation may be realized if the material is strained, or
grown on a substrate \cite{footnote2}.
It also occurs in \srrho,
the isostructural and isoelectronic $4d$ counterpart of \sriro.
As expected for $4d$-orbitals, the bandwidth is smaller. 
%Also the SOC is smaller ($\zeta_{SO}$=$0.161$\,eV), 
%as can be seen in the right panel of Fig.~\ref{fig1}, 
%e.g.~from the band splitting %between the
%at $\Gamma$.
As before, we construct Wannier $j_{\textrm{eff}}$ orbitals 
from the bands within the energy window $[-2.67,0.37]$\,eV;
the Wannier $j_{\textrm{eff}}$ character along the bands of \srrho\ is
shown on the right hand side of Fig.~\ref{fig1}.
Again, there is a mixture of the \jeff12\ and 
\j32\ $|m_j|$=$3/2$ orbital character along the four 
bands crossing the Fermi level. However, since 
the SOC is weaker in this compound ($\zeta_{SO}$=$0.161$\,eV)
than in its $5d$-counterpart, the effective 
splitting between the bands \jeff12\ and \j32\ $|m_j|$=$3/2$ is smaller
and these two states have similar filling
($n_{1/2}$=$1.42$, $n_{3/2,|3/2|}$=$1.64$), whereas the
\j32\ $|m_j|$=$1/2$ state is filled ($n_{3/2,|1/2|}$=$1.96$).
The system is thus close to a $3/4$-filled two-band model, which 
favours a metallic state.

Moreover, the Hubbard interactions are smaller than in \sriro:
a cRPA calculation as above gives 
$F^0$=$1.6$\,eV and $J$=$0.3$\,eV for \srrho.
Indeed, the weaker hybridization
of the Rh-$4d$ states with the O-$2p$ 
locates the latter about $1$\,eV
higher in energy than in \sriro.
As a result,
the Coulomb interactions are screened more efficiently
in \srrho\ compared to \sriro.
By computing the interaction within cRPA for an 
artificial \srrho\ system in which we shift
the O-$p$-states down by $1$\,eV 
we verify that $F^0$ is indeed increased.

Our \LDADMFT\ calculations, Fig.~\ref{fig4}, are in 
good agreement with energy 
distribution curves obtained by \ARPES\ at $10$\,K 
\cite{BaumbergerPRL96-2006,PerryNJP8-2006}. We oberve  
a large electron-like pocket of radius $0.65-0.69$\,\AA$^{-1}$ 
and a smaller hole pocket of radius $0.19-0.18$\,\AA$^{-1}$ 
in the undistorted Brillouin zone.
Further structures are found between -$0.1$ and -$0.2$\,eV 
along $\Gamma$-$X$ and around -$0.1$\,eV along $\Gamma$-$M$. 
The orbitally-resolved spectral functions allow to attribute 
both of these structures to the \j32\ $|m_j|$=$1/2$ orbital, whereas 
the hole-like $\alpha$-pocket around $\Gamma$ is 
of \j32\ $|m_j|$=$3/2$ type. The two other pockets, 
$\beta_M$ and $\beta_X$, are (mostly) of type \jeff12. 

We find that \srrho\ is close to the Mott
transition (which would occur for interaction strengths of  
$F^0$=$1.8$-$2.0$~eV while keeping $J$ fixed).
The occupancies of the \jeff12\ and \j32\ $|m_j|$=$3/2$ spin-orbitals 
are $0.63$ and $0.89$, resp. 
The quasi-particle weights for \jeff12\ and \j32\ $|m_j|$=$3/2$
have been estimated as $Z_{{1}/{2}}$=$0.5$ and 
$Z_{{3}/{2},|{3}/{2}|}$=$0.6$. 

In conclusion, we have shown that only the cooperative effect
of SOC, lattice distortions and Coulomb correlations
drives the 5$d$ compound \sriro\ insulating, due to a complete
spin-orbital polarization resulting in an effective 
half-filled one-orbital (more precisely, two-spin-orbital)
system.
The isostructural and isoelectronic 4$d$ compound \srrho\ 
has smaller SOC and Coulomb interactions,
leading to a less dramatic reduction of spin-orbital
fluctuations and, as a result,  
a two-orbital (or four spin-orbital) metal.
The structural and electronic similarities between \sriro\
and La$_2$CuO$_4$ -- both are Mott insulators with one hole
in an effective one-orbital system -- together with the
pronounced difference in their magnetic structure, 
suggest that the transport properties of {\it doped} \sriro\ 
may give valuable information about the role of magnetic
fluctuations in high-T$_c$ superconductivity.
Oxygen-deficient \sriro\ is not superconducting
\cite{KornetaPRB82-2010}.
Our electronic structure calculations show, however, 
that doping by impurities with extremely weak SOC
should act as giant perturbations to the spin-orbital
structure, in analogy to introducing a {\it magnetic}
impurity into a system with weak SOC. 
It would be most interesting to introduce carriers
into the system while conserving as much as possible the
very specific electronic structure, e.g. 
by doping with heavy atoms such as Os or Re.

%\acknowledgements
This work was supported by the French Agence Nationale de la
Recherche under project CorrelMat and GENCI/IDRIS under project 20111393.
We acknowledge useful discussions with R. Arita, J.-S. Bernier,
M. Ferrero, M. Imada, J. Kunes, J. Mravlje, O. Parcollet, and H. Takagi.

%\bibliography{sr2iro4-article_trunc}

\begin{thebibliography}{29}
\bibitem[{foo({\natexlab{a}})}]{footnote1}
\bibinfo{note}{Note however that for computational reasons these parameters
  were calculated for the undistorted phases without SOC. Screening with \eg\
  states is excluded in the range of the hybridisation gap in order to mimick
  the band structure of the distorted material. We then transform to the basis
  of t$_{2g}$-projected total angular momentum eigenstates, keeping density-density
  interaction terms only.}
\bibitem[{foo({\natexlab{b}})}]{footnote2}
\bibinfo{note}{Due to the similarity of the lattice constant of SrTiO$_3$,
  \sriro\ on SrTiO$_3$, e.g. should be nearly undistorted.}
\end{thebibliography}

\end{document}